\documentclass[nologo,a4paper,10pt]{ETHpaper}
\usepackage[british]{babel}
\usepackage{blkarray} 
\usepackage[square,numbers]{natbib}

\newcommand{\edge}[2]{e_{#1\to #2}}
\renewcommand{\sign}[2]{s_{#1\to #2}}
\newcommand{\G}[1]{\mathcal G_{#1}}
\renewcommand{\S}[1]{\mathcal S_{#1}}
\def\ppp/{{\footnotesize ($+++$)}}
\def\ppm/{{\footnotesize ($++-$)}}
\def\pmm/{{\footnotesize ($+--$)}}
\def\mmm/{{\footnotesize ($---$)}}

\title{Reconstructing signed relations from interaction data}

\titlealternative{Signed Relations from Interaction Data}

\author{Georges Andres\footnote{geandres@ethz.ch, corresponding author}, Giona Casiraghi\footnote{gcasiraghi@ethz.ch}, Giacomo Vaccario\footnote{gvaccario@ethz.ch}, Frank Schweitzer\footnote{fschweitzer@ethz.ch}}
\authoralternative{G. Andres, G. Casiraghi, G. Vaccario, F. Schweitzer}
\address{ETH Z\"urich, Chair of Systems Design\\ Weinbergstrasse 56/58,  Z\"urich, Switzerland}
\www{\url{http://www.sg.ethz.ch}}
\reference{}

\begin{document}
\maketitle

\abstract{
Positive and negative relations play an essential role in human behavior and shape the communities we live in.
Despite their importance, data about signed relations is rare and commonly gathered through surveys.
Interaction data is more abundant, for instance, in the form of proximity or communication data.
So far, though, it could not be utilized to detect signed relations.
In this paper, we show how the underlying signed relations can be extracted with such data.
Employing a statistical network approach, we construct networks of signed relations in four communities.
We then show that these relations correspond to the ones reported in surveys.
Additionally, the inferred relations allow us to study the homophily of individuals with respect to gender, religious beliefs, and financial backgrounds.
We evaluate the importance of triads in the signed network to study group cohesion.

\smallskip
\textbf{\emph{Keywords:} Signed Networks, Interaction Data, Homophily}
}

\section{Introduction}

Social interactions and signed relations are distinct yet related facets of human behavior.
Social interactions are short-lived contacts during which individuals exercise directed or reciprocal influence over one another~\citep{schmid_reciprocity2021}.
Individuals can interact via different means, and their interactions may repeatedly occur over time. 
Signed relations, such as friendship and enmity, are interpersonal relations characterized by a \emph{sign} (positive or negative) reflecting how one person feels or thinks about another.
Signed relations are long-lived and change less frequently as more effort is required to form or change them.

While social interactions and signed relations are different, they are coupled to each other--relations acting as drivers for interactions.
A positive relation commonly induces more interactions, while a negative one hinders them~\citep{homan1950FOFJustification}.
Moreover, humans perceive surrounding patterns of positive and negative relations~\citep{freeman1988human}, to which they adapt~\citep{heider_1958}.
Over time, such adaptations can lead to interactions appearing mostly within cohesive groups, potentially leading to echo-chambers~.
Negative links may be formed across opposing groups, pushing communities towards segregation and, eventually, to polarization~\citep{groeber_2014,schweighofer_2020_2}.

To understand such phenomena quantitatively, we require data on the positive and negative relations, which is rare.
Interaction data is the more abundant alternative.
However, they do not directly inform us about the relations among individuals.
This leads to the problem of inferring meaningful information only from interaction data.
Usually, this problem is addressed by taking the network perspective, where nodes represent individuals and edges their interactions~\citep{guimera2009missing, peixoto2014hierarchical, newman2018networkstructure, coscia2017network, Scholtes2017}.
Network filtering \citep{radicchi2011information} and backboning methods \citep{serrano2009extracting} can extract relevant connections from observed noisy interactions and find successful applications in biology~\citep{wang2010process,mora2018identifying} and economics~\citep{glattfelder2009backbone}.
Alternative methods use thresholding rules \citep{wuchty2011ThresholdingInt} or take a topic modeling perspective~\citep{tumminello2005tool}.
All these methods, though, can at most be applied to the study of \emph{unsigned} relations.
For the recovery of \emph{signed} relations, we require novel approaches.

We introduce a statistical network method to infer weighted signed relations from a collection of unsigned, repeated interactions.
We will refer to it as the $\Phi$-method.
It relies on the main assumption that a statistical \emph{over}-representation of interactions signals a \emph{positive} relation and an \emph{under}-representation signals a \emph{negative relation}~\citep{Nanumyan2018}.
This assumption is motivated by the longstanding theoretical argument that individuals with positive relations are more likely to interact~\citep{rapoport1954triadicClosure,homan1950FOFJustification} and its empirical evidence across different communities~\citep{jones_postivieTieStrength_2013,pappalardoTieStrength_2012,urenaTieStrength_2020}.
Moreover, the idea that negative relation induces fewer interactions is supported by the arguments that people avoid individuals who are considered a source of discomfort rather than pleasure~\citep{harrigan_avoidanceNegativeTies2017,labiance_avoidanceNegativeTies2006}.

To demonstrate our $\Phi$-method, we utilize four classical interaction datasets of social communities.
These are a karate club in a university~\citep{zachary_1977} (KC), a windsurfer community~\citep{freeman1988human} (WS), a high school in France~\citep{dataset_FrenchHighschool} (HS) and participants in the Nethealth project~\citep{dataset_NetHealth} (NH).
These social communities are chosen because they, in addition to interactions, contain information about social relations that can be used to validate our method.

With our method, we reconstruct the underlying relational networks of the four communities.
The inferred signed relations allow us to study pairs and triads of individuals in a new light.
We illustrate the strength of having access to the complete relational structure of communities, which we represent using a weighted signed network.
To this end, we investigate the pairwise homophily, relational triads, and cohesiveness of groups in the communities.
Note that we refer to social communities (KC, WS, HS, NH) rather than to those detected by community-detection algorithms.

\section{Results}
\label{sec: results}

\begin{figure}[h]
  \centering
  \includegraphics[width=\textwidth]{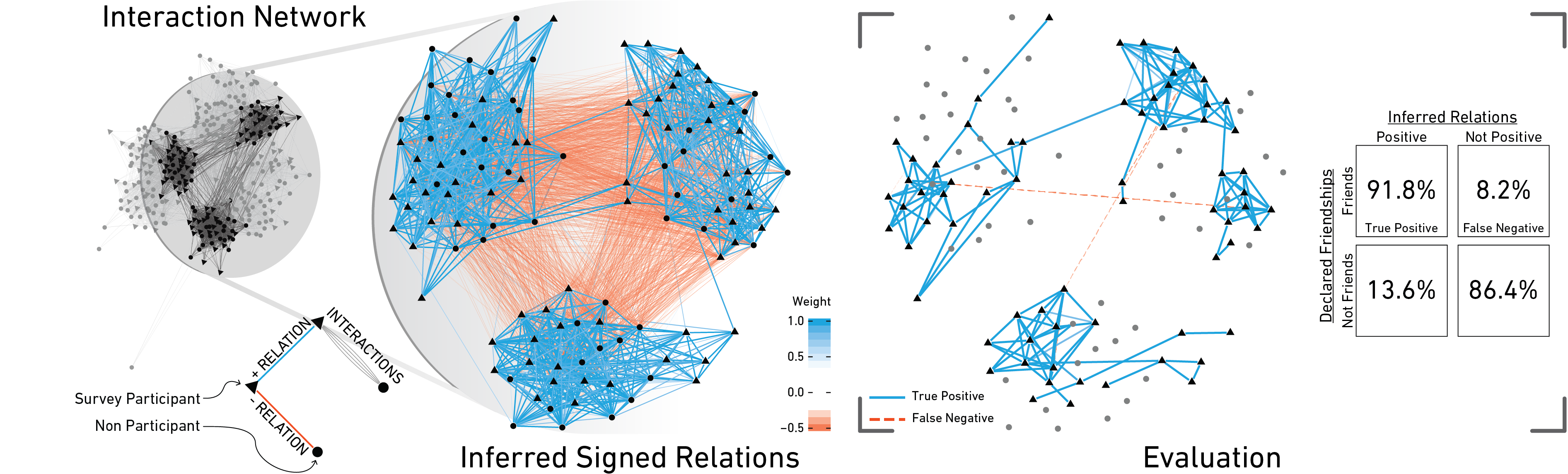}
  \caption{
  \textbf{(left)} Interaction network $\mathcal{G}_{HS}$ from the HS dataset. Nodes represent individuals and edges recorded interactions between them. Multiple interactions are shown by parallel edges.
  \textbf{(center)} Inferred signed network $\mathcal{S}_{HS}$ shown only for a subset individuals. Positive relations are represented by blue edges (darker colour refers to larger weight).
  \textbf{(right)} Network of declared friendship relations among individuals.
  We report a summary of the evaluation in a confusion matrix.
  }
  \label{fig:method}
\end{figure}

\paragraph{Inference of signed networks}
To infer the weighted signed networks $\S{i}$ for the four communities KC, HS, WS and NH (extended details provided in Methods), we first construct an interaction network $\G{i}$.
An edge $\edge{v}{w}$ in $\G{i}$ is created every time an interaction between individuals $v$ and $w$ is observed in the respective dataset.
Furthermore, each dataset contains a small set of \emph{reported relations} obtained by directly surveying a subset of the individuals.
Such reported relations are either binary (i.e., positive or not positive) or continuous (i.e., how strong they are).

In~\cref{fig:method}, we visualize the interaction network $\G{\text{HS}}$ only for HS, which records interactions between students in a French high-school divided into 9 classes.
From $\G{\text{HS}}$ we infer the weighted signed network $\S{\text{HS}}$.
In $\S{\text{HS}}$, we observe clusters of positive relations with weak negative ties between the clusters.
This pattern matches the class separation within the high-school.
If we compare $\S{\text{HS}}$ to the declared friendships provided in the survey (\cref{fig:method} (right)), we see that most declared friendships are within classes and only few across classes.

To obtain the weight and the sign $\sign{v}{w}$ of the links in $\S{\text{HS}}$, we use the $\Phi$-method.
For each pair $(v, w)$ of individuals, the weight of the relation $\sign{v}{w}$ is obtained as a linear combination of the probability that two individuals are interacting more than expected with the probability of interacting less than expected (see Methods for details).
The coefficients of this linear combination are estimated based on the few reported relations in the community.
Once determined, this allows us to infer both positive and negative relations between \textit{all} individuals, going beyond previous approaches~\citep{tang_negativeLinkPrediction2015}.

\paragraph{Accurate prediction of reported relations}
Using the $\Phi$-method, we accurately predict the reported relations between individuals.
To evaluate this accuracy, we perform both an in-sample and an out-of-sample prediction task where the dependent variable is the reported relation and the predictor the value of $\sign{v}{w}$.
We detail the results of the prediction tasks in \cref{tab:Quality of the models}.
For HS, NH, and KC, the reported signed relations are categorical (individuals being friends or not, or individuals feeling a strong, weak or no relation at all).
Hence, we evaluate $\S{i}$ by means of standard classification methods and list the resulting sensitivity, specificity, and balanced accuracy (see Methods).
All these scores are remarkably high and above $80\%$, which holds for both the in-sample and the out-of-sample predictions.
For WS, the reported signed relations are continuous.
Thus, we model them with a linear regression.
We evaluate the goodness of fit using the R$^{2}$ and the root-mean-squared-error.
These continuous relations are harder to model, as they were obtained through a convoluted interview process.
Our goodness of fit suffers from this with an R$^{2}$ just above $0.3$.

We find that the $\Phi$-method is robust in handling unseen data.
For the HS and NH dataset, we preserve a very similar accuracy between the in-sample and the out-of-sample prediction, the same holds for the difference in R$^{2}$ in the WS dataset.
The most considerable accuracy loss occurs in the case of the small KC dataset where the specific train-test split has a significant impact.
In the supplementary material, we further show that the $\Phi$-method outperforms other approaches for predicting relations based on thresholding rules or network modularity.

\begin{table}[ht]
  \small
\begin{center}
\begin{tabular}{l c c c | c}
& HS & NH & KC & WS \\
  \hline
  \textbf{Model specification} & friends $\sim$ $\phi$ &  friends $\sim$ $\phi$ & faction $\sim$ $\phi$& closeness $\sim$ $\phi$\\
  &&&&\\
  \textbf{In-sample} &&&&\\
  Sensitivity   & $0.831$ & $0.831$ & $1$ &  \\
  Specificity   & $0.931$ &  $0.985$ & $0.938$  &   \\
  Balanced Accuracy & $0.881$  & $0.908$ & $0.969$ &  \\
  R$^{2}$   &   & &  & $0.307$ \\
  RMSE   &   & &  & $0.118$ \\
  &&&&\\
  \textbf{Out-of-sample} &&&&\\
  Sensitivity   & $0.8$ & $0.821$ & $0.875$ &  \\
  Specificity   & $0.941$ &  $0.985$ & $0.875$  &   \\
  Balanced Accuracy & $0.871$  & $0.903$ & $0.875$ &  \\
  R$^{2}$   &   & &  & $0.302$ \\
  RMSE   &   & &  & $0.120$ \\
\end{tabular}
\caption{Quality of the model for in-sample and out-of-sample predictions. We report the sensitivity, specificity, and balanced accuracy for the binary HS, NH, and KC. For the continuous relations in WS, we report the R$^{2}$ and the root-mean-squared-error (RMSE). Overall, the model quality is good for the binary relations and worse for the continuous ones. The model is robust as the out-of-sample prediction only loses little compared to the in-sample prediction.}
\label{tab:Quality of the models}
\end{center}
\end{table}

\begin{figure}[h]
  \centering
  \begin{subfigure}{0.24\textwidth}\centering
  	\textbf{Gender (HS)}\\
  	\includegraphics[width=\textwidth]{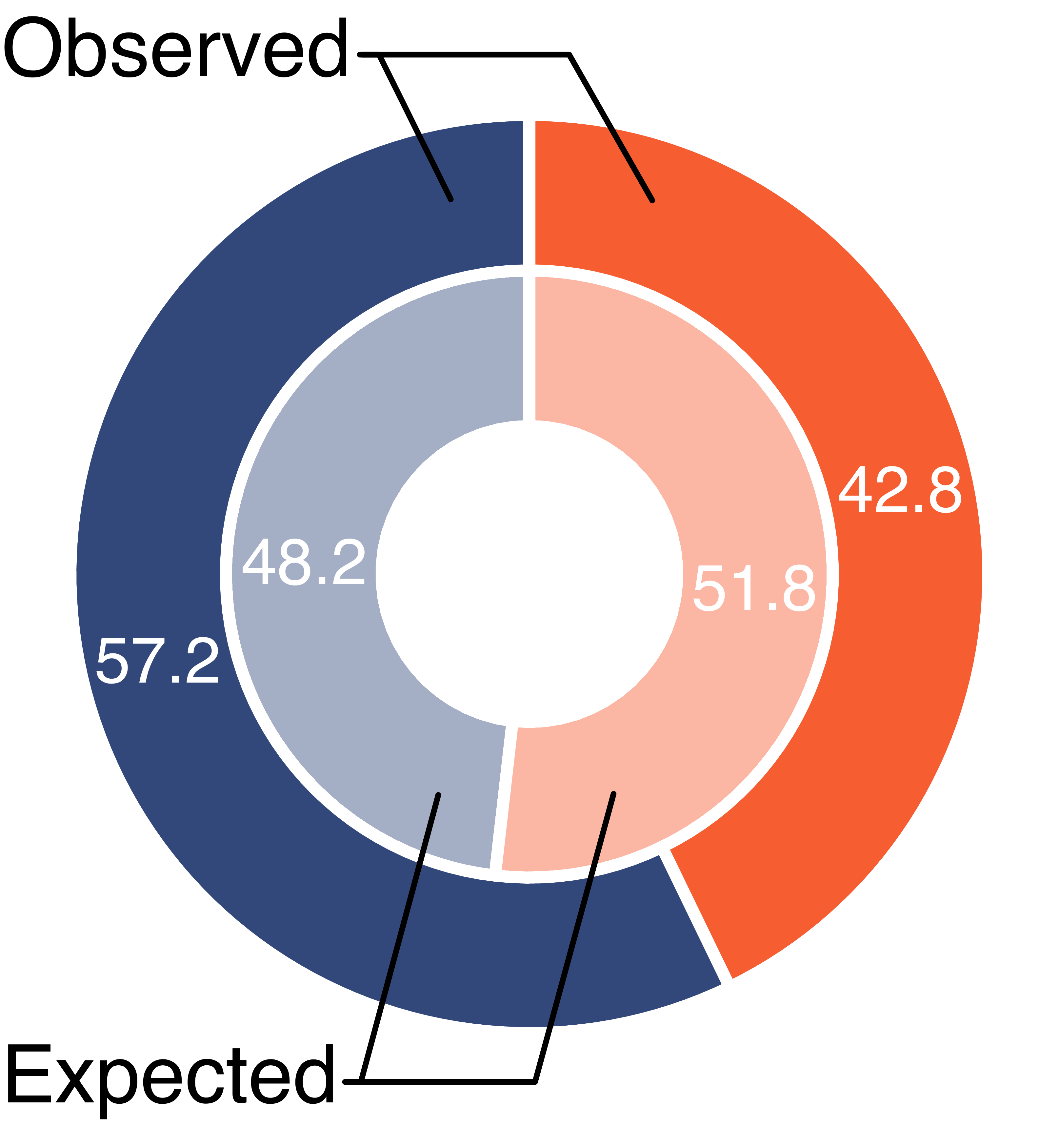}
  \end{subfigure}
  \begin{subfigure}{0.24\textwidth}\centering
  	\textbf{Gender (NH)}\\
  	\includegraphics[width=\textwidth]{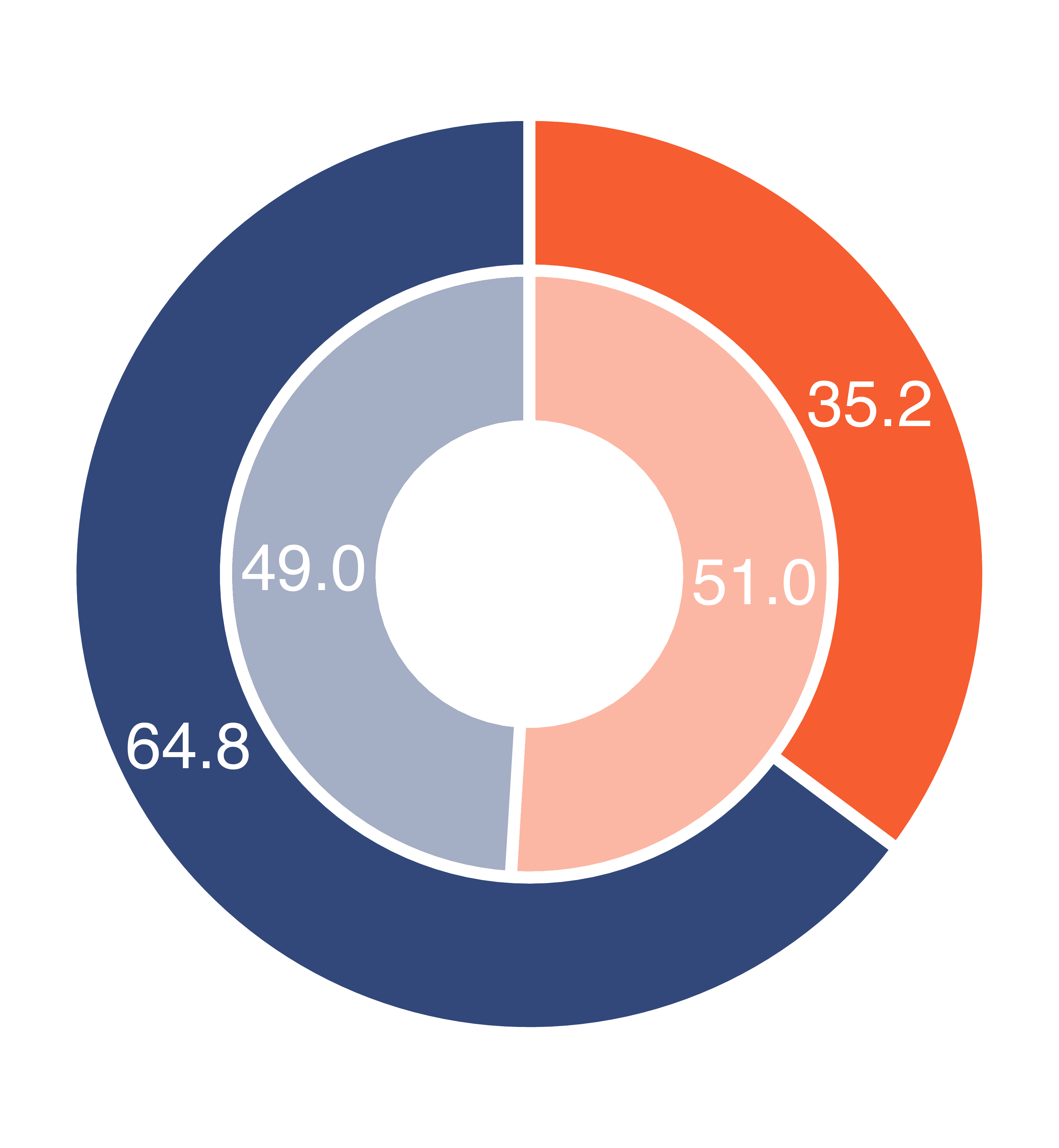}
  \end{subfigure}
  \vline
  \begin{subfigure}{0.24\textwidth}\centering
  	\textbf{Religion (NH)}\\
  	\includegraphics[width=\textwidth]{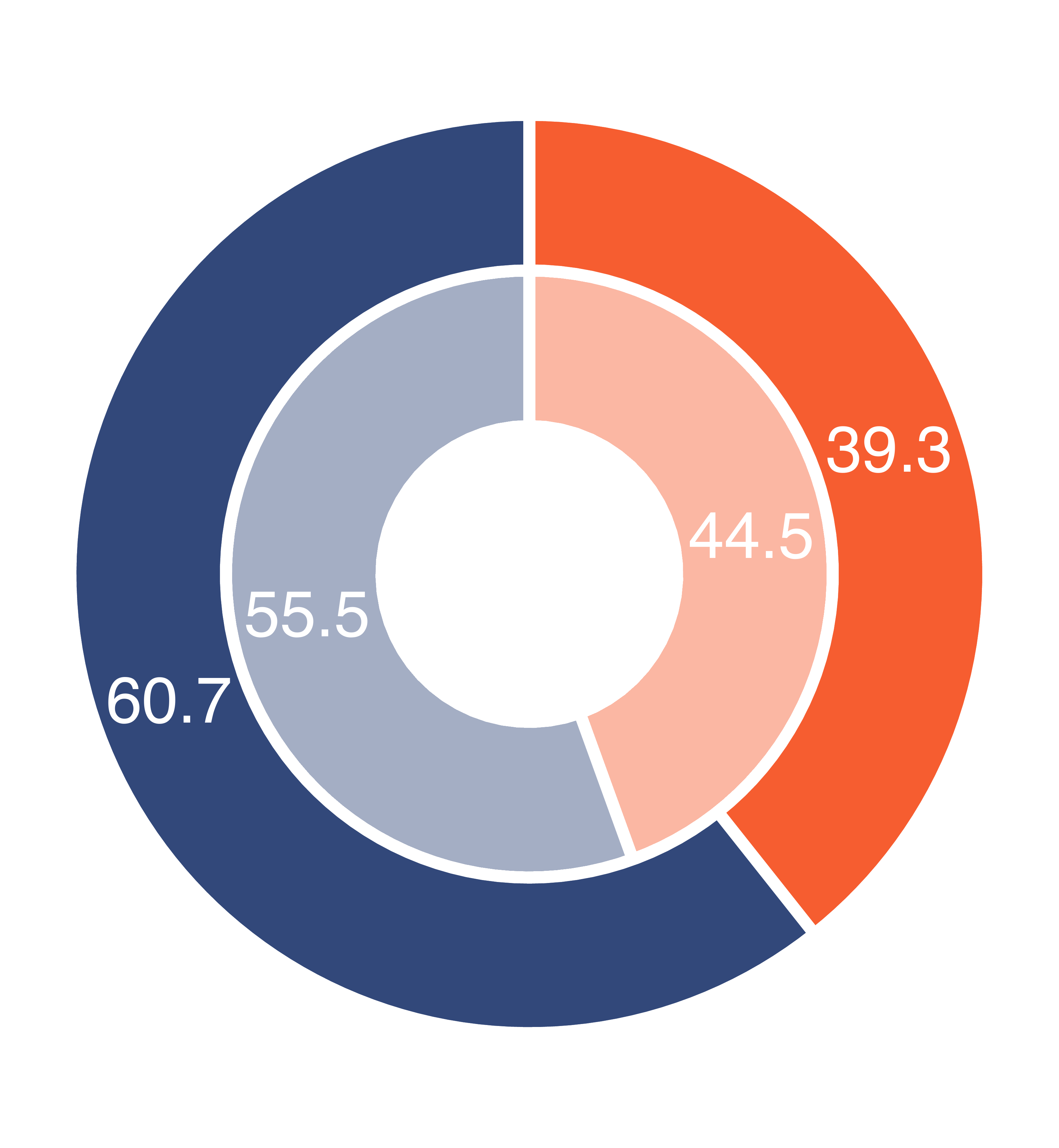}
  \end{subfigure}
  \begin{subfigure}{0.24\textwidth}\centering
  	\textbf{Income (NH)}\\
  	\includegraphics[width=\textwidth]{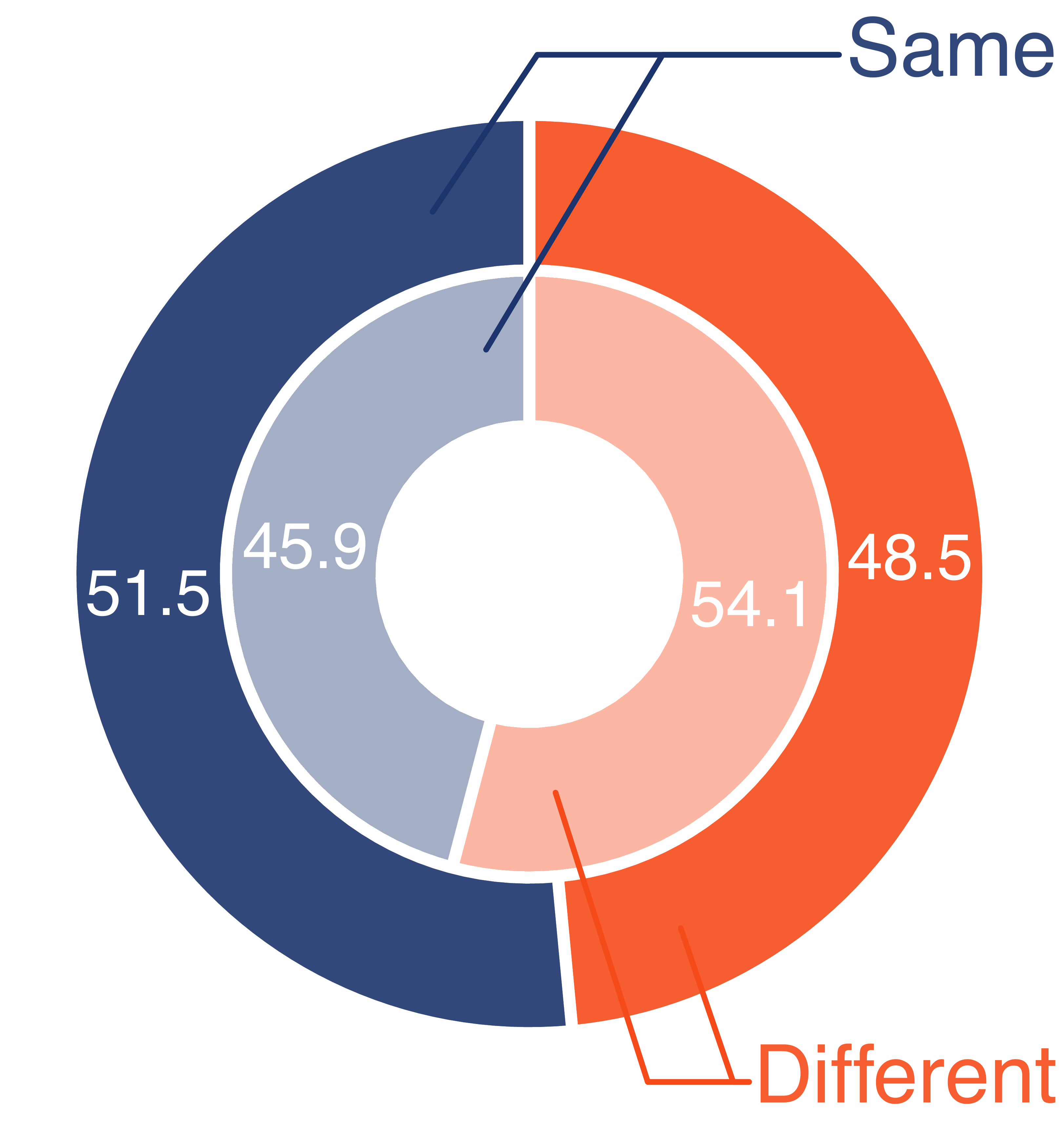}
  \end{subfigure}
  \caption{
    (Left) Gender homophily in HS and NH.
    (Right) Religion and income homophily in NH.
    The outer ring shows the probability (in percentage) that individuals with a positive relation also have the same gender, relgion or parental income.
    The inner circle refers to the random sampling.
    While all three types of homophily are present, gender homophily is the strongest.}
  \label{fig:homophily}
\end{figure}

\paragraph{Homophily}
Homophily is the phenomenon of similar individuals being more likely to form positive relations.
In the inferred signed networks $\S{\text{HS}}$ and $\S{\text{NH}}$, we find strong gender homophily, i.e., the specific case in which similarity is defined by gender.
To test the presence of this phenomenon, we compare two probabilities (in percentage): i) the probability that individuals with a positive relation also have the same gender and ii) the probability that randomly sampled pairs of individuals have the same gender.
These are shown in~\cref{fig:homophily} in the i) outer and ii) inner circles.
We only have data about genders in the NS and HS datasets, so we restrict the analysis to these two datasets.
We find that the probability that individuals with a positive relation also are of the same gender is larger compared to the reference probability of randomly sampled pairs being of the same gender (\cref{fig:homophily}).
Precisely, compared to the reference case, it is approximately $20\%$ and $30\%$ more likely that individuals with a positive relation have the same gender in the HS and NH dataset, respectively.
By performing a binomial test, we verify that these results are statistically significant (see Methods for details).

Apart from gender, we find that religion and parental income homophily are of lesser importance to university students.
This is shown in~\cref{fig:homophily}, by comparing 64.8 vs 49.0 for gender to 60.7 vs 55.5 for religion and 51.5 vs 45.9 for parental income.
Only for this dataset do we have such additional information.
The probability that friends have similar religious beliefs or parental income is slightly larger than in the reference case, but nevertheless significant.

\paragraph{Beyond dyadic properties}
Thanks to our analysis, we have attributed a signed relation to each pair of individuals.
The datasets contain additional information about the belonging of these individuals to different groups (e.g. classes, memberships).
By looking at triads composed of three individuals, we can now characterize these groups.
Considering only the sign of relations, four types of triads $T_{\tau}$ can appear:  \ppp/ ($T_{1}$), \ppm/ ($T_{2}$), \pmm/ ($T_{3}$), \mmm/ ($T_{4}$).
For each triad $t=(v,w,z)$ of a given type $T_{\tau}$, we assign a weight $\omega_{t}$ by multiplying the weighted signs $\sign{v}{w}$, $\sign{w}{z}$, and $\sign{z}{v}$~\citep{schweitzer_Iching2022}.
We define group \emph{cohesion} by means of triads $T_{1}$ with three positive relations \ppp/.
Group \emph{conflict} on the other hand, is defined by those triads $T_{2}$ that have one negative link \ppm/.

Through the weights of the triads, we can quantify the importance of each type of triads for groups (see Methods for details).
We can distinguish formal groups (e.g. classes) from informal groups, for example the two groups in KC centered around the leaders JA and HI.
Analyzing the networks of signed relations $\mathcal{S}_{HS}$, $\mathcal{S}_{KC}$ and $\mathcal{S}_{WS}$, we find that cohesion strongly outweighs conflict only in HS, which contains formal groups.
Differently, informal groups emerging in WS and KC show weaker cohesion and a higher presence of conflict.
Specifically,~\cref{tab:triads} shows, that \ppp/ ($T_{1}$) triads have high importance within the groups of HS ($0.98$ and $0.96$).
In the informal groups of WS and KC, their importance decreases up to $0.45$.
Moreover, in the JA group of KC, conflict has as much importance than cohesion.
Across all analyzed communities, the importance of relational triads with many negative relations, \pmm/ ($T_{3}$) and \mmm/ ($T_{4}$), is marginal.

Our analysis of KC further highlights leaders' influence on group formation.
While, at the time of the data collection, KC consisted of a single community, it eventually split into two groups centered around two leaders, JA and HI~\citep{zachary_1977}.
Analyzing these two groups separately, we find that the triads \emph{involving} their leaders are strongly cohesive:
\ppp/ ($T_{1}$) triads involving HI and JA have an importance of 0.72 and 0.59, respectively (see \cref{tab:triads} for details).
However, when considering triads \emph{not involving} the leaders, we only find cohesion in HI's group (0.63).
JA's group instead is dominated by conflict (0.54).
Hence, we have revealed that the presence of the influential leader is the major characteristic defining the group.

\begin{table}[ht]\centering
  \small\hspace{.1cm}\hfill
  \begin{tabular}{l c | c | c }
    & \textbf{WS} & \textbf{HS} & \textbf{KC}\\
    \hline
      &  \textbf{G1} $\vert$ \textbf{G2} &  \textbf{C1} $\vert$ \textbf{C2} &  \textbf{HI} $\vert$ \textbf{JA}\\
    {\footnotesize $+++$} &  $\pmb{0.73}$  $\vert$ $\pmb{0.83}$ & $\pmb{0.98}$  $\vert$ $\pmb{0.96}$ & $\pmb{0.68}$ $\vert$ $\pmb{0.45}$ \\
    {\footnotesize $++-$} & $0.23$  $\vert$ $0.15$ & $0.02$ $\vert$ $0.03$ & $0.24$ $\vert$ $\pmb{0.45}$  \\
  \end{tabular}
\hfill\vline\vline\hfill
   \begin{tabular}{l c c}
    & \includegraphics[width=0.4in]{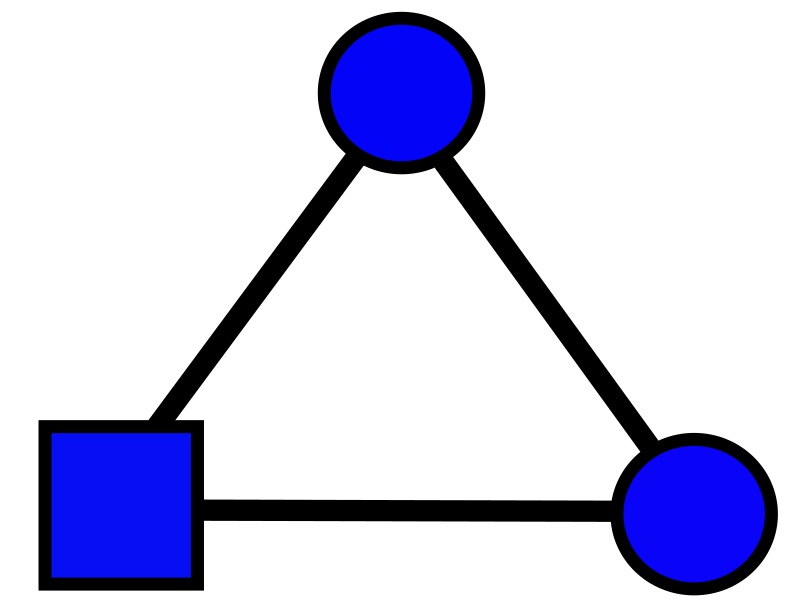} & \includegraphics[width=0.4in]{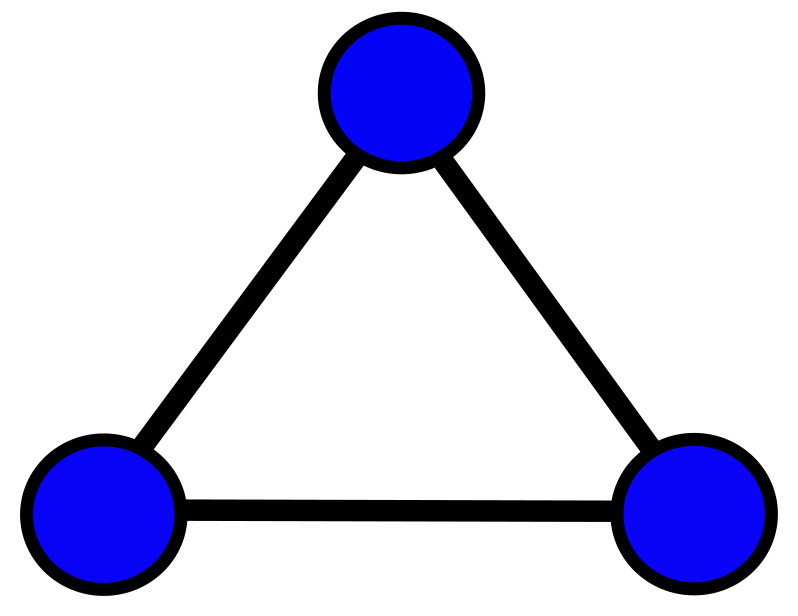} \\
    & \textbf{HI} $\vert$ \textbf{JA} & \textbf{HI} $\vert$ \textbf{JA} \\
     {\footnotesize $+++$}  & $\pmb{0.72}$ $\vert$ $\pmb{0.59}$  &  $\pmb{0.63}$ $\vert$ $0.28$\\
     {\footnotesize $++-$}   & $0.28$ $\vert$ $0.38$ &  $0.19$ $\vert$ $\pmb{0.54}$
\end{tabular}
\hfill\hspace{.1cm}
\caption{(Top) Importance of triad types \ppp/ and \ppm/ for different communities.
  Each community features groups and the importance of the triads is calculated within these groups.
  In all groups but the one of John A. (JA) in KC, the importance of cohesion outweighs conflict.
  (Bottom) Left are triads in KC involving the leaders of the groups (squared node), right triads not involving the leaders.
Mr. Hi's group is always characterized by cohesion, while John A.'s shows mostly conflict when he is not present.}
\label{tab:triads}
\end{table}

\section{Discussion}

Our work contributes to the study of human relations by unlocking data sources previously not usable for such investigations.
To infer signed relations between individuals, we have employed data about face-to-face contacts (HS), SMS and phone calls (NH), proximity (WS) and co-attendance (KC).
Traditionally, weighted signed relations are obtained with surveys, an expensive and hardly scalable approach.
Instead, interaction data is abundantly available.
Despite the different types of data, we have shown that our methodology is well suited to extract signed relations.
Therefore, social scientists, behavioral researchers, and psychologists can now use interaction data in new ways.

Our central assumption is that positive relations imply more and negative relations fewer interactions.
This way of linking interactions to relations is a long-standing assumption in social science~\citep{homan1950FOFJustification}, which has been widely tested for positive relations~\citep{jones_postivieTieStrength_2013,pappalardoTieStrength_2012,urenaTieStrength_2020}.
In the case of negative relations, instead, it has rarely been explored, mainly due to a lack of data.
The $\Phi$-method fills this gap.

Our broader perspective allows quantifying social phenomena such as homophily, cohesion, and conflict within groups.
For instance, we have confirmed that gender homophily is essential in establishing positive relations, such as friendship.
Additionally, we have found that leaders can strongly influence the cohesion of a group.
This result can be related to the theories of social status and structural balance, according to which individuals adapt their behavior in response to their surroundings~\citep{weberStatus_1919,ridgewayStatus_2006,heider_1958,cartwright_1956}.

Finally, the ability to infer signed relations from interaction data enables to study how relations \emph{evolve} over time.
Social theories about structural balance, status, or social impact postulate different mechanisms for \emph{relational changes}.
We can now test these mechanisms by leveraging the fine-grained temporal resolution of interaction data.
This opportunity paves the way for future research to explore the evolution of signed relations and their effect on communities with an unprecedented resolution.

\section{Data}

We require data about social communities containing both interactions \emph{and} declared relations, gathered through surveys.
While such data is, in general, scarcely available, we leverage four datasets fulfilling our requirements. 
They vary in size, number and type of interactions, and form of surveyed relations.
We summarize this information in~\cref{tab: Data summary}.

The data ranges from small communities of under $50$ individuals to larger ones encompassing hundreds of people.
In these datasets, an interaction $\edge{v}{w}$ indicates proximity between, colocation, or communication events through phone calls, SMS, and WhatsApp between two individuals $v$ and $w$.
In the two datasets HS and NH, interactions were collected automatedly.
Thus, they feature the most interactions: up to roughly $2 \cdot 10^{6}$ for NH.
In the other two datasets, instead, interactions were recorded manually by researchers.
The surveyed relations $r_{vw}$ either indicate a quasi-continuous closeness, belonging to one of four factions, or a binary friendship, i.e., people being friends or not.

\begin{table}[ht]\centering
  \small
  \begin{tabular}{lccccc}
         \hline
        & Nodes & Interactions & Relations &  Interaction Type & Relation Type\\
        \hline
  \textbf{WS}  & $43$    & $1206$         & $903$     & Proximity & Closeness\\
  \textbf{KC} & $34$    & $231$          & $30$      & Co-attendance & Faction belonging\\
  \textbf{HS} & $327$   & $67\,613$         & $406$     & Face-to-Face Proximity & Friendship\\
  \textbf{NH}     & $698$    & $1\,987\,527$        & $1353$     &  Communication & Friendship\\
  \hline
  \end{tabular}
  \caption{Summary of the main features of the data.}
  \label{tab: Data summary}
   \end{table}

\paragraph{Windsurfer (WS)}

\begin{figure}[ht]
  \centering
    \textbf{(a)}\includegraphics[width= 0.4\textwidth]{figures/windsurfers_weighted_interactions}
    \textbf{(b)}\includegraphics[width= 0.4\textwidth]{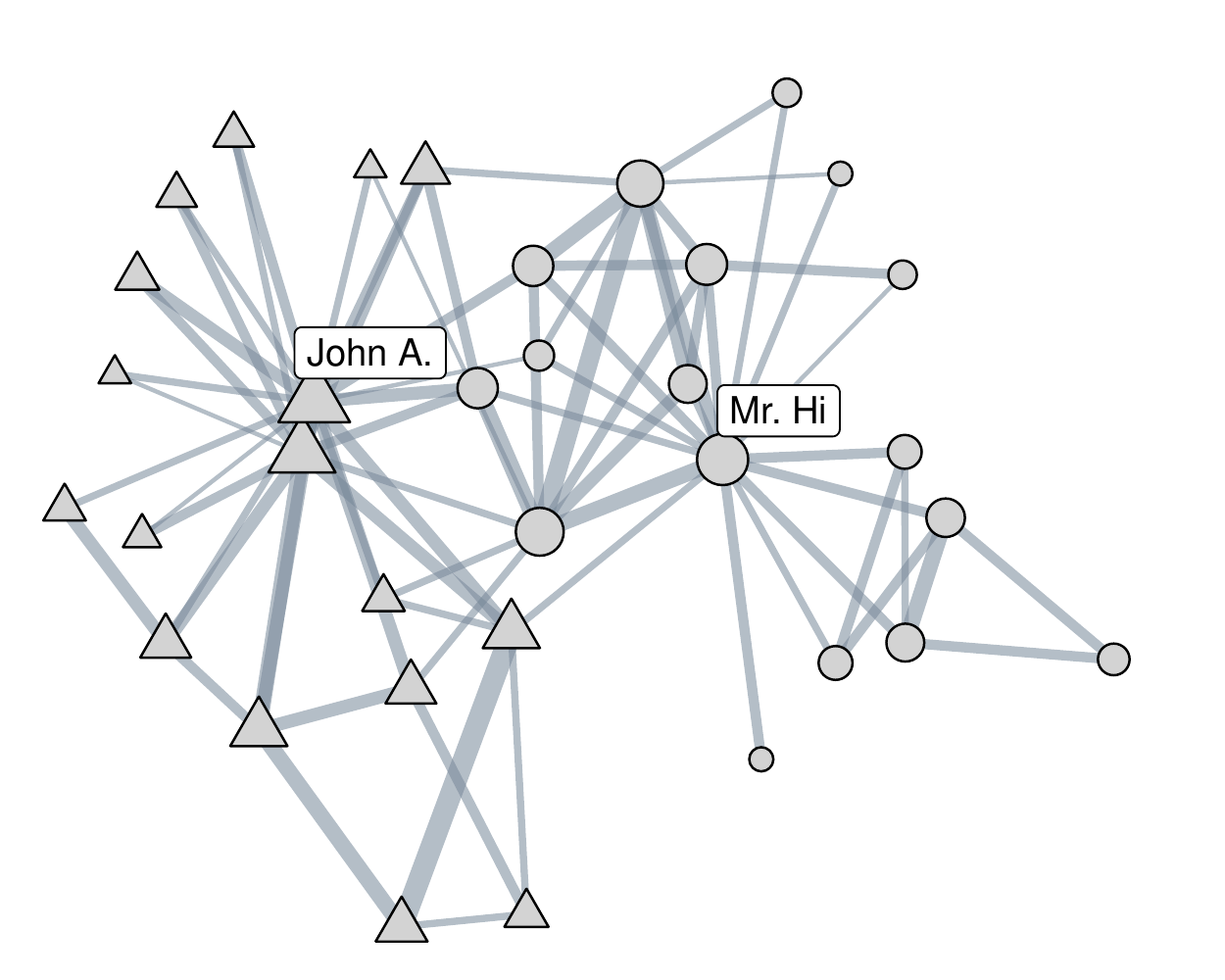}
    \textbf{(c)}\includegraphics[width= 0.4\textwidth]{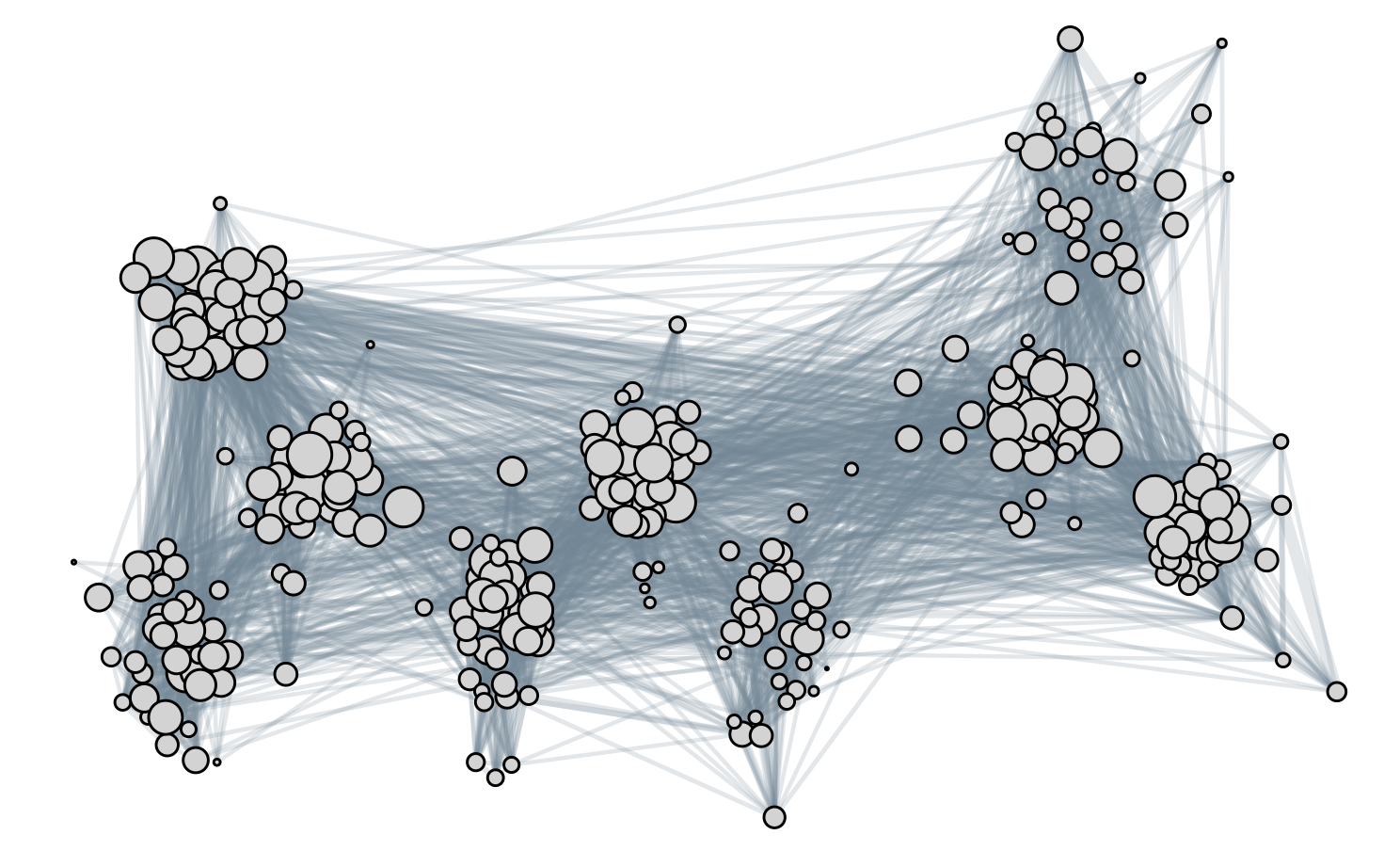}
    \textbf{(d)}\includegraphics[width= 0.4\textwidth]{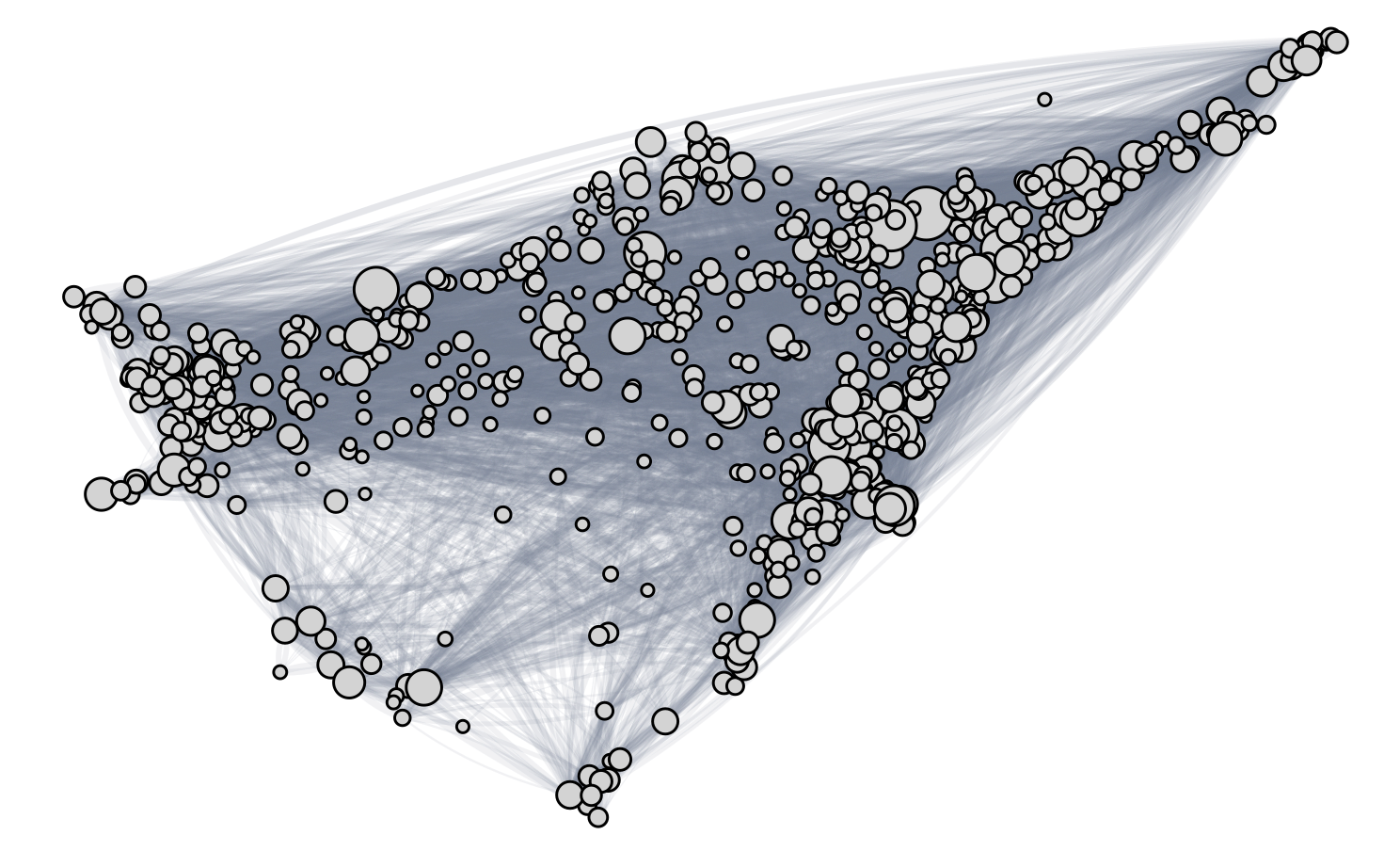}
    \caption{Interaction networks visualized for (a) WS, (b) the KC, (c) HS and (d) NH. Link weights in the figures are proportional to interaction counts.}
    \label{fig:interaction networks}
\end{figure}

The study of the windsurfer community took place in California in the fall of 1986, with the authors being long-time members of this community~\citep{freeman1988human}.
The windsurfers were naturally dividing themselves into two groups, newcomers and older members, but there was no display of intergroup conflict.
They were observed over 31 days, each day for two 30 min intervals.
The interactions can loosely be defined as proximity events, people sitting together for lunch, or social exchanges.
Looking at the interaction network (\cref{fig:interaction networks}a) makes it clear that most interactions took place within the two informal groups.
All community members were interviewed shortly after the conclusion of the observation period.
They were asked to perform a sorting task to identify how close they were to each other.
This closeness is rescaled to a number in $(0,1)$  and represents the relations in this dataset.
Even though the authors describe a dataset of $54$ surfers, only data about 43 of them was released.

\paragraph{Zachary's Karate Club (KC)}
This dataset contains interactions between 34 members of a university karate club over three years.
The recorded interactions occurred not during the karate lesson but in different contexts.
Like the windsurfer community, the karate club had two factions that ``were never organisationally crystallized'' and ``[...] not named''\citep{zachary_1977}.
However, the factions had two leaders the club president (John. A.) and the karate instructor (Mr. Hi).
These factions arose due to a dispute between the leaders over an increase in the costs of lessons.
At a certain point, the club split into two clubs, one led by John. A. and the other by Mr. Hi.
The club members mainly chose the leader they wanted to join according to the factions they were in before the split~\citep{zachary_1977}.
The interaction network  (\cref{fig:interaction networks}b) makes these factions visible before the split, while inter-faction contacts are still present.
Before the split, club members were asked which faction they saw themselves in and whether that sentiment was strong or weak.
These declarations form the relations in our analysis.
The data also contains information about each member's final group after the split.

\paragraph{French Highschool (HS)}
As a third community, we consider a high school in France.
\citep{dataset_FrenchHighschool} have recorded face-to-face interactions between students from four programs and organized them into nine classes.
This was done using RFID trackers, which only trigger when individuals are close and facing each other.
The interactions are recorded while being at school over five days.
Interactions are mainly concentrated within classes, which becomes apparent when considering the network visualization (\cref{fig:interaction networks}c).
Nevertheless, students interacted with alters from other classes, possibly during breaks.
On top of the interactions, information was collected about positive social relations, i.e., friendship.
Unfortunately, no information about negative relations was collected.

\paragraph{Nethealth Project (NH)}
Lastly, we studied the Nethealth Project, a long-lasting (2015-2019) study conducted by the Center for Network Science and Data at the University of Notre Dame~\citep{dataset_NetHealth}.
It investigates the social networks and health of initially around $700$ undergraduate students, comprising pair-wise interaction data as well as responses to surveys administered in $8$ waves over the study period.
Interactions were recorded through communication events in the form of in- and out-going calls and messages from the participants' phones.
We construct the interaction network  (\cref{fig:interaction networks}d) only including people who have at some point participated in the study and have given their consent to the use of their data.
The sheer size of the interaction network does not allow us to extract much information from its visualization.
However, we see that the degrees of the nodes vary greatly, between $0$ at least and $89950$ at most.
The data contains surveyed friendships, which constitute the relations we use in our work.
As there were multiple `waves' of surveys, in our analysis, we focus on one wave, namely the second one.
This wave contains the most individuals, as subsequently there were some drop-outs.
We then only consider interactions happening between the first and second surveys.
Our results remain stable over the other waves.

\section{Methods}
\label{sec:materials}
\subsection{Inferring signed relations}

\paragraph{The $\Phi$-method}
The $\Phi$-method relies on the central assumption that over/under-representations of interactions signal positive/negative relations, a longstanding hypothesis in social sciences \citep{homan1950FOFJustification}.
To quantify these over-and under-representations, we compare the observed interaction counts between individuals to a network null model, the hypergeometric ensemble of random graphs (HypE)~\citep{casiraghi2021}.
By employing a network null model, we define an expectation for the number of interactions between individuals.
This expectation should account for all factors that bias the observed number of interactions beyond the effect of signed relations~\cite{Scholtes2017}.
In this work, we specifically account for the heterogeneity in the activities of the different individuals.
That means, we account for the fact that a very active individual is more likely to interact with others irrespectively of whether they share a positive or negative relation.
Similarly to a standard configuration model~\cite{fosdick_2018}, HypE allows explicit modeling of such heterogenous activities and enables the estimation of network- and dyadic- sampling probabilities through closed form expressions~\cite{casiraghi2021}.
It does so by modeling the network generation as a sampling process without replacement from a carefully designed urn.

The urn is filled with a given number of balls, each representing a possible directed edge between two nodes $v$ and $w$.
An edge $\edge{v}{w}$ from $v$ to $w$ is considered to be in this set of possible edges if the nodes have non-zero in- and out-degrees $k^{out}_{v}$ and $k^{in}_{w}$, respectively.
To account for the different levels of activity of different individuals, we specify the maximum number $\Xi_{vw}$ of possible edges between each pair of individuals to be proportional to the the activity---i.e., degree---of each individual in the network.
To do so, we define a matrix $\pmb{\Xi}$, whose entries $\Xi_{vw}$ are given by $k^{out}_{v}k^{in}_{w}$.
It directly follows that $\sum_{vw} \Xi_{vw} = m^{2}$ is the total number of possible edges, and thus the number of balls in the urn.
A network realization $\pmb{X}$ with $m$ edges is given by sampling $m$ balls from this urn without replacement.
This sampling procedure is akin to hypergeometric sampling, and the probability of finding the observed network configuration $\pmb{A}$ is given by:
\begin{equation}
  \label{eq:ensemble proba}
  \Pr\left( \pmb{X} = \pmb{A}\right) = \frac{\prod_{vw}\binom{\Xi_{vw}}{A_{vw}}}{\binom{m^2}{m}}.
\end{equation}
\Cref{eq:ensemble proba} defines HypE, the network ensemble that we use to estimate the pair-wise over-and under-representation of interactions.
This ensemble has the benefits of incorporating interdependencies between pairs of individuals, preserving individuals' activity and attractiveness, and being analytically tractable.
For more details, we refer to~\cite{casiraghi2021}.
While in this work, we focus only on incorporating the \emph{activity} of individuals into our null model, it is in principle possible to extend the null model to account for more complex factors, e.g., block or sub-group structures~\cite{casiraghi2019block}.
However, these extensions are beyond the scope of this article.

From \cref{eq:ensemble proba}, we extract the two marginal probabilities $P(X_{vw}<A_{vw})$ and $P(X_{vw}>A_{vw})$, where $A_{vw}$ is the observed number of interaction between $v$ and $w$ and $X_{vw}$ is an hypergeometric random variable:
\begin{align}
  \label{eq:cummulative marginals}
  \Pr\left( X_{vw} < A_{vw}\right) &= \sum_{a_{vw}=0}^{A_{vw}-1}\frac{\binom{\Xi_{vw}}{a_{vw}} \binom{M - \Xi_{vw}}{m - a_{vw}}}{\binom{M}{m}}\\
  \Pr\left( X_{vw} > A_{vw}\right) &= \sum_{a_{vw}=A_{vw}+1}^{\Xi_{vw}}\frac{\binom{\Xi_{vw}}{a_{vw}} \binom{M - \Xi_{vw}}{m - a_{vw}}}{\binom{M}{m}}
\end{align}
Intuitively, when the first probability is high, it is unlikely to find as many interactions as we observed, indicating an over-representation~\cite{Scholtes2017,larock2020hypa} and, therefore, a positive relation.
The same reasoning holds for the second probability, indicating a negative relation.
Extending the approach of \citep{Nanumyan2018}, we construct the signed relations by taking the difference of these probabilities, weighted according to some constants in what we call the $\Phi$-method $\mathcal{M}_\Phi$:
\begin{equation}
  \label{eq:phi}
  \phi_{vw}(a,b) = aP(X_{vw}<A_{vw}) + b P(X_{vw}>A_{vw})
\end{equation}
As shown in the following, we can learn the community-dependent constants $a$ and $b$ when we have access to data about the relations between a small number of individuals in the community.
When this data is not available, we assume a symmetric influence of over- and under-representation, i.e. $a=-b=1$.

\paragraph{Constructing the signed networks: training on data}

Whenever we have access to data about interactions \emph{and} relations between \emph{some} individuals, we can train the $\Phi$-method to find optimal parameters $\hat a$ and $\hat b$ to infer signed relations.
By extrapolating the learned parameters to \emph{all} pairs in the community, we compute \cref{eq:phi} and construct full signed networks from only a few reported relations.

We employ simple machine learning techniques to estimate the parameters in \cref{eq:phi}.
Our aim is to classify the reported relation $r_{vw}$ based on the value of $\phi_{vw}(a,b)$:
\begin{equation}\label{eq:machine}
  r_{vw} \sim \phi_{vw}(a,b) + c\,.
\end{equation}
Whenever we have binary relations, e.g., $r_{vw} \in \{\text{Friend}, \text{Not Friend}\}$, we perform the classification in \cref{eq:machine} by means of a logistic regressions.
In the case of continuous relations, e.g., $r_{vw}$ refers to some `closeness'  $\in (0,1)$, we use linear regressions.
If multiple categories are possible, e.g., $r_{vw} \in \{\text{Friend}, \text{Positive Attitude}, \text{Neutral}, \text{Negative Attitude}, \text{Enemy}\}$, multinomial or cumulative link methods~\citep{agresti_categoricalData_2002} are employed, depending on whether the categories are ordered or not.

The classification just described gives us estimates $\hat a$ and $\hat b$ for the parameters in \cref{eq:phi}, obtained for the subset of individuals for which reported relations $r_{vw}$ exist.
With these, we can extrapolate our findings to the whole community, generating the signed network $\S{}$, whose links $\sign{v}{w} = \phi_{vw}(\hat a,\hat b)$.
In~\cref{tab: coefficients}, we report the coefficients estimated for all datasets.
These coefficients are community-dependent.
However, $a$ is always positive, and $b$ is always negative.
This finding demonstrates that having a high over-representation in interactions increases the probability of having a surveyed friendship.
Similarly, having a high under-representation decreases this probability.
Additionally, the only dataset with a large negative $\hat b$ is KC.
This community is also the only one in which a known conflict arose.
For the other communities, $\hat b$ tends to be small in absolute value, giving weakly negative relations.

The coefficient $c$ in \cref{eq:machine}, provides a baseline from which the value of $\phi_{vw}(a,b)$ can be related to the reported relations.
Thus, we do not employ such value in constructing the signed network $\S{}$.

\begin{table}[h]
  \small
\begin{center}
\begin{tabular}{c c | c c c c}

coefficient & predictor & WS & KC & HS & NH \\
\hline
  $\hat a$ & $P(X_{vw}<A_{vw})$ & $0.19$ & $2.42$ & $4.71$ & $5.67$\\
  $\hat b$ & $P(X_{vw}>A_{vw})$ & $-0.06$ & $-0.92$ & $-0.21$ & $-0.64$\\
\end{tabular}
\caption{Estimated coefficients $\hat a$ and $\hat b$ for over- and under-representation for the four datasets studied. $\lvert\hat b\rvert$ is always smaller than $\lvert\hat a\rvert$ for all datasets, indicating the presence of weak negative links. Only for KC we have a large negative coefficient. This is expected as it is the only community in which a known conflict emerged.}
\label{tab: coefficients}
\end{center}
\end{table}

\paragraph{Comparing $\Phi$ to other methods}

In the following, we show that the $\Phi$ method outperforms two other methods used to infer relations.
The first one is a threshold method $\mathcal{M}_{T}$.
The user defines a threshold on the interactions over which individuals are assumed to be friends.
Similarly, they are assumed to be enemies below this threshold.
We assume one threshold for all pairs in the community and this threshold can be learned from the known relations.
Specifically, we use as a predictor the interaction counts $A_{vw}$ in the regression methods:
\begin{equation}
  r_{vw} \sim \alpha A_{vw} + c\,.
\end{equation}
This method disregards any heterogeneities in the individuals, their different levels of activity in the community, or their popularity.
We can partly alleviate this by factoring in the degrees of the individuals when defining their relations.
By quantifying the expected number of interactions between two individuals based on their degrees, we reach a formulation akin to the one used in the well-known network modularity~\citep{newmanModularity_2003,leichtDirectedModularity_2008}.
We call this model the modularity method $\mathcal{M}_{M}$.
Formally, it can be written as follows (for directed networks):
\begin{equation}
\mu_{vw} = A_{vw} - \frac{k_{v}^{out}k_{w}^{in}}{m}
\end{equation}
In the undirected case, total degrees are substituted $k_{v}^{out}=k_{v}$ and $k_{w}^{in}=k_{w}$ and the right-hand side is divided by two.
While the modularity method now partly accounts for heterogeneities, it disregards that the two individuals we study are part of a larger system, namely the whole network.
To compare it to the $\Phi$-method, we use this $\mu_{vw}$ as a predictor in the regression to learn appropriate scaling parameters.

Below, we demonstrate that our proposed $\Phi$-method outperforms both the threshold and the modularity methods in identifying the known relations.
To do so, we perform cross-validation on a training subset of the data and validate the learned representations of the relations on a separate testing subset.
This out-of-sample prediction task tests the different methods' ability to predict relations in unseen data based on its learned specification.

In~\cref{tab:Model comparison}, we report our findings for all datasets.
For the three datasets with categorical relations (HS, NH, KC), we are interested in correctly identifying the known relations, i.e., the true positives and true negatives.
Additionally, we are dealing with unbalanced data, where most pairs have no relation.
Therefore, we report the balanced accuracy (BA) score, the mean of sensitivity and specificity, which fits our problem best.
We report the R$^2$ coefficient for the continuous relations in WS.
Consistently across all but KC, the $\Phi$-method outperforms the other two methods.
In KC, a highly modular graph, the modularity method performs similarly to the $\Phi$-method.

\begin{table}[ht]
  \small
\begin{center}
  \begin{tabular}{l c c c | c}
    & HS (BA) & NH (BA)& KC (BA)& WS (R$^{2}$)\\
    \hline
   $\mathcal{M}_{T}$   & $0.813$ & $0.870$ & $0.750$ & $0.179$ \\
   $\mathcal{M}_{M}$  & $0.824$ &  $0.860$ & $0.875$  &  $0.244$ \\
    $\mathcal{M}_{\Phi}$   & $0.871$  & $0.903$ & $0.875$ & $0.302$ \\
\end{tabular}
\caption{Comparing $\Phi$ to other models. Balanced accuracy/ R$^{2}$ from in-sample and out-of-sample prediction through cross validating the inferred relations in all four datasets. While the in-sample comparison remains inconclusive, the out-of-sample sees a drastic improvement of the $\Phi$-method over the other two.}
\label{tab:Model comparison}
\end{center}
\end{table}

\subsection{Statistical evaluation of homophily}

To evaluate the statistical significance of our results on homophily for NH and HS, we perform a binomial test.
Let $m_\text{SG}$ be the number of pairs that share the same gender and $m_\text{DG}$ the number of opposite pairs.
The probability to randomly sample a pair with the same gender from the full data is then $p=m_\text{SG}/(m_\text{SG}+m_\text{DG})$.
If we have $n$ friends in total and $l$ friends who also share the same gender (success), the p-value of the binomial test is given by:

\begin{equation}
  \label{eq:p-val}
  p = P(Y \geq k) = \sum_{i = l}^{n} \binom{n}{i} p^{i}(1-p)^{n-i}
\end{equation}
where $Y$ is a random variable.
If this probability is low, it is improbable to observe at random as many or more homophilous friends as we do in the data.
For the HS, we find a p-value of $p_\text{HS}^\text{G}=1.6\cdot10^{-6}$.
For NH, the p-values are $p_\text{NH}^\text{G} = 3.16\cdot10^{-95}$, $p_\text{NH}^\text{I} = 1.67\cdot10^{-6}$ and $p_\text{NH}^\text{R} = 3.70\cdot10^{-5}$ for gender, income and religion respectively.
All p-values are significant ($<0.05$).

\subsection{Quantifying the importance of triads}

Let $T_{\tau= \{1,2,3,4\}}$ be the set of all triads of either one of the four types: \ppp/, \ppm/, \pmm/, \mmm/.
We quantify the importance of a given triad type $T_{\tau}$ as:

\begin{equation}
  \label{eq:importance}
               n(T_{\tau}) = \sum_{t \in (T_{\tau})} \omega_{t} = \sum_{t \in (T_{\tau})} \|\phi_{vw\in t}\| \cdot \|\phi_{wz\in t}\| \cdot \|\phi_{zv\in t}\|
\end{equation}
The sum runs over all triads $t$ is the set $T_{\tau}$.
The subscript $vw\in t$ signifies that the link between $v$ and $w$ is in the triad $t$.
Note that we use the absolute value of the $\Phi$-measure.
Thus, we consider the weight of the relation when evaluating the importance of a given triad.
This way, triads containing mainly weak links will contribute less to the importance.

To obtain a number comparable across communities, we normalize the importance of each triad type over the total importance of all triad types.
\begin{equation}
  \label{eq:relative importance}
               I({T_{\tau}}) = \frac{n(T_{\tau})}{N}
\end{equation}
where $N = n_{(+++)} + n_{(++-)} + n_{(+--)} + n_{(---)}$.
Such a normalization gives us the relative importance, which is the number we report for the different datasets in \cref{tab:triads} in the main text.

\small \setlength{\bibsep}{1pt}

\end{document}